# Structure of the Universe: An Approach Based on Gravitational Energy
Anthony Baldocchi

Observations and data collected through the years, from microwave background to the Hubble constant have constantly upturned modern physical theories and have led to speculation and the creation of new ideas. Often the unexplainable data is explained away through the creation of some previously unknown and physically irrelevant idea. Several examples of this are dark matter and the "false vacuum," both used to explain the expansion of the universe. Neither of these ideas has a solid basis in physics or in collected data. They are simply creations used to fill in the holes in existing theories, and are no more valid, though easily more plausible, than a giant fourth dimensional tortoise spewing matter into our universe through a hole at the center.

A viable explanation should be able to be determined from supportable ideas and basic properties. Using certain basic assumptions that have, thus far, not been disproved and thus may possibly be true, the expansion is explainable. Understanding of the model presented here is most easily understood, although does not require, the assumption that space-time is compressed or curved around matter. This is caused by, rather than the cause of, gravity. Gravity cannot be an innate property because gravity does work and if gravity were simply a property it would violate the Third Law of Thermodynamics. In order to preserve this law, the energy imparted by gravity must be transferred from the body causing gravity. But even an uncharged object at rest causes gravity, so the source of the energy appears to be nonexistent. The only possible source of gravitational energy is the matter itself. Relativity has shown a correlation between matter and energy. Gravity could thus be caused by a direct conversion of matter to energy, this energy causing an attractive gravitational field. Thus, the longer matter exists, the more energy is given off, and the less matter remains. Energy is given off in quanta, but the packets exhibited through gravity are smaller than the energy contained in the currently considered fundamental particles of matter. This does not, however, falsify this theory or cause a requisite of smaller particles of matter. Instead, the fundamental particles lose mass as they release energy. At first glance, it would seem that such an effect would be easily detectable. However, equations prove otherwise. Force divided by mass equals acceleration. An object losing mass would, it would seem have to accelerate unjustifiably. However, if the gravitational body were also losing mass, the force it exhibits would also be decreasing. Therefore, force divided by mass would equal the original acceleration.

    All matter releasing energy compresses the space around it proportional to its mass. However, as gravitational energy is expelled by the gravitational body, its mass decreases by the same ratio and releases the compressed, or flattens the curved, space=time, causing an expansion around any gravitational body. With space=time expanding around every gravitational

body, the same expansion rate would be observed in every direction, and the Hubble constant is easily accounted for, as there is approximately an equal distribution of galaxies, so there would be approximately an equal expansion of space out to a specific distance in each direction, as the expansion due to each body between the observation point and the radius distance would be the same net value in every direction.

How does the origin of the universe fit into this model?  Currently accepted ideas hold that the universe began at a single point, followed by a period of rapid expansion, followed by a period of regular expansion.  According to this model, all matter would be compressed to a point, or exist in an infinitely curved space.  When this point began to uncompress, the incredible density of matter would have caused much overlapped expansion; that is, the radius of expansion overlaps many other radii of expansion, causing a much higher rate of expansion at this time. Eventually, though, matter is so dispersed that the greater part of the expansion radius does not overlap a significant amount of the matter in the universe, so the rate of expansion shrinks to that of a single unit of matter, unit being defined as a locally joined body of many particles that acts as a single body; the Earth, for instance, is not considered to be the sum of atoms and the gravitational force the sum of that of each atom, but rather it is considered a unit.

This model has not yet accounted for the creation of galaxies and the grouping of matter. There are two considerations here. The first is that units of matter existed as units before the expansion began, that is when the universe was infinitely compressed, and thus once the rate of expansion was reduced to the level of a single unit of matter, these units attracted other matter and eventually formed the stars and galaxies. However, this idea requires the assumption that the matter was clumped to begin with, which is entirely unjustifiable and smacks of "divine intervention." The other possibility is that matter did not escape the radius of great acceleration at the same time, but rather the outside slowed first, colliding with matter that is recently escaping the greatest expansion, and as the "expansion field," the radius of great expansion, shrinks de to the escape of matter from said field, "pockets" of matter result, which eventually become the galaxies and universe as we know it.

What will happen at the end of such a universe? As the universe flattens out, matter loses mass and energy. Eventually, matter will run out of energy, the universe will flatten out, and everything, save energy and space-time, will no longer exist. Long before this time, life will have ceased to exist as blood ceases to flow; cells disintegrate as they don't have enough energy to hold the various constituents of matter together, so the universe will become an equal distribution of cold energy, small enough that there isn't enough energy localized to form even the most fundamental constituent of matter. The universe of matter and life will have "melted" into a puddle of energy thinly layered on top of the maximum plane of space-time. Beyond this, it is impossible to reason or measure what will happen, but it could be reasonably assumed that space and time themselves will separate, lacking anything to hold them together. At such low energies, there is little to hold the bonds of space and time and the universe, as we or anyone knows it would unravel until there is nothing left.